\documentclass[12pt]{article}
\usepackage[round,comma,numbers,sort&compress]{natbib}
\usepackage{graphicx}
\usepackage{amssymb}
\usepackage{xspace}
\usepackage{times}

\newcommand{\refjnl}[1]{{\em #1\ }}
\newcommand\aj{\refjnl{Astron J.}}%
\newcommand\anndap{\refjnl{Annales d'Astrophysiques}}%
\newcommand\araa{\refjnl{Annu. Rev. Astron. Astrophys.}}%
\newcommand\apj{\refjnl{Astrophys. J.}}%
\newcommand\apjl{\refjnl{Astrophys. J.}}%
\newcommand\apjs{\refjnl{Astrophys. J. Suppl. Ser.}}%
\newcommand\apss{\refjnl{Astrophys. Space Sci.}}%
\newcommand\aap{\refjnl{Astron. Astrophys.}}%
\newcommand\mnras{\refjnl{Mon. Not. R. Astron. Soc.}}%
\newcommand\pasp{\refjnl{Publ. Astron. Soc. Pac.}}%
\newcommand\nat{\refjnl{Nature}}%
\newcommand\etal{{\em et al.\xspace}}

\newenvironment{sciabstract}{%
\begin{quote} \bf}
{\end{quote}}

\newcounter{lastnote}
\newenvironment{scilastnote}{%
\setcounter{lastnote}{\value{NAT@ctr}}%
\addtocounter{lastnote}{+1}%
\begin{list}%
{\arabic{lastnote}.}
{\setlength{\leftmargin}{.22in}}
{\setlength{\labelsep}{.5em}}}
{\end{list}}

\title{Massive-Star Supernovae as Major Dust Factories}

\author{
}

\date{}

\newcommand\sun{\odot}
\newcommand\farcs{\mbox{$.\!\!^{\prime\prime}$}}

\newcommand\arcsec{\mbox{$^{\prime\prime}$}\xspace}%
\newcommand\micron{\mbox{$\mu$m}\xspace}%
\newcommand{\um}{\micron}
\newcommand{\ujy}{$\mu$Jy\xspace}

\newcommand{\kms}{${\rm km\;s}^{-1}$\xspace}
\newcommand{\msun}{$M_\sun$\xspace}

\newcommand{\lsun}{$L_\sun$\xspace}

\renewcommand{\micron}{$\mu$m\xspace}

\newcommand{\SST}{{\em SST}\xspace}

\newcommand\ion[2]{#1$\;${\small\rmfamily{#2}}\relax}%

\begin{document}


\maketitle

\vspace{-5\baselineskip}
\begin{center}
Ben E. K. Sugerman$^{1\ast}$, 
Barbara Ercolano$^2$,
M.\ J.\ Barlow$^2$, 
A. G. G. M. Tielens$^3$,
Geoffrey C.\ Clayton$^4$,
Albert A. Zijlstra$^5$
Margaret Meixner$^1$,
Angela Speck$^6$,
Tim M. Gledhill$^7$,
Nino Panagia$^1$,
Martin Cohen$^8$,
Karl D.\ Gordon$^9$,
Martin Meyer$^1$,
Joanna Fabbri$^2$,
Janet E. Bowey$^2$,
Douglas L.\ Welch$^{10}$,
Michael W.\ Regan$^1$,
Robert C.\ Kennicutt, Jr.$^{11}$
\\
\ \\
{$^{1}$Space Telescope Science Institute, 3700 San Martin
  Dr., Baltimore, MD 21218, USA}\\
{$^2$Department of Physics and Astronomy, University College London, Gower 
  Street, London WC1E 6BT, UK}\\
{$^3$Kapteyn Astronomical Institute, P.O. Box 800, 9700 AV Groningen,
  Netherlands}\\
{$^4$Dept.~of Physics \& Astronomy, Louisiana State
  University, Baton Rouge, LA 70803, USA}\\
{$^5$School of Physics and Astronomy, University of Manchester,
  P.O. Box 88, Manchester M60 1QD, UK}\\
{$^6$Dept. of Physics \& Astronomy, University of Missouri,
 316 Physics, Columbia, MO 65211, USA}\\
{$^7$Dept.\ of Physics, Astronomy, and Maths, University of
  Hertfordshire, College Lane, Hatfield AL10 9AB, UK}\\ 
{$^8$Monterey Institute for Research in Astronomy, 200 Eighth Street,
  Marina, CA 93933, USA}\\ 
{$^9$Steward Observatory, University of Arizona, 933 North Cherry
Avenue, Tucson, AZ 85721, USA} \\
{$^{10}$Dept.~of Physics and Astronomy, McMaster University,
  Hamilton, Ontario L8S 4M1, Canada}\\
{$^{11}$Institute of Astronomy, University of Cambridge, Madingley
  Road, Cambridge, CB3 0HA, UK}\\
\ \\
\normalsize{$^\ast$To whom correspondence should be addressed; E-mail:
sugerman@stsci.edu.}
\end{center}




\begin{sciabstract} 
We present late-time optical and mid-infrared observations of the
Type-II supernova 2003gd in NGC 628.  Mid-infrared excesses consistent
with cooling dust in the ejecta are observed 499-678 days after
outburst, and are accompanied by increasing optical extinction and
growing asymmetries in the emission-line profiles.  Radiative-transfer
models show that up to 0.02 solar masses of dust has formed within the
ejecta, beginning as early as 250 days after outburst.  These
observations show that dust formation in
supernova ejecta can be efficient and that massive-star supernovae can
be major dust producers throughout the history of the Universe.
\end{sciabstract}


\section*{Introduction}


Millimeter observations of high-redshift ($z>6$) quasars 
have revealed the presence of copious amounts of dust when the
Universe was as young as 700 million years \cite{Ber03}.
At the present day, dust in the interstellar medium of the Milky Way
and other galaxies is generally thought to be injected mainly by the
gentle winds of low mass stars when they evolve onto the Asymptotic
Giant Branch \citep{Tie98}. However, stellar-evolution timescales of
these low-to-intermediate mass stars are too long for them to be a
major contributor to the dust budget in the early universe
\cite{Dwe98}. Instead, dust in the early universe must reflect the
contribution from rapidly-evolving (1-10 million years) massive stars
which return their nuclear ashes in explosive type II supernova
(SN) events. Theoretical studies have long suggested that dust can
condense in the ejecta from core collapse (e.g.\ type II) SNe
\cite{Cer65} and calculations predict condensation of 0.08--1 \msun of
dust within a few years, depending on metallicity and progenitor mass
\cite{Dwe88,Koz89,TF01}. There is also evidence for the
origin of some dust in type II SNe based on isotopic composition of
stardust isolated in meteorites \cite{Cla97}.


Direct observational evidence for efficient dust formation in SN
ejecta is, however, lacking, largely because SN explosions are rare
and far apart.
Dust formation was detected in the ejecta of SNe 1987A and 1999em, but
only some $10^{-4}$ \msun were inferred for each
\cite{Luc91,Woo93,Elm03}, a factor up to $10^3$ smaller than typical
SNe would have had to produce in order to contribute efficiently to
the early-Universe dust budget \cite{ME03}.  Similarly low dust masses
have been measured in evolving SN remnants using the recently-launched
Spitzer Space Telescope \cite{Hin04}, however its mid-infrared (IR)
instruments are most sensitive to warm (50--500 K) material, while
dust in these remnants has almost certainly cooled to $<30$ K.  Cold
dust has been detected in remnants in the far-IR and sub-mm, however
such observations risk strong contamination by cold, unrelated dust
clouds along the line of sight \cite{Kra04}.  As such, the best way to
demonstrate dust condensation in SN ejecta is to study them within a
few years of their explosion, during the epoch of condensation when
the ejecta are much hotter than interstellar dust.  The high
sensitivity of the Spitzer's mid-IR detectors allows us to sample very
young core collapse SNe within $\sim$20 Mpc and opens up the whole
nearby Universe for such studies.  Here we report on such a study of
the type II-P SN 2003gd in the galaxy NGC 628. A rare combination of
contemporaneous optical and mid-IR observations of this well-studied
SN with a known stellar progenitor mass of $8^{+4}_{-2}$ \msun
\cite{VDy03,Sma04} provides an excellent test case for the efficiency
of dust formation in SN ejecta.


\section*{Data in Support of Dust Production}

As dust condenses in SN ejecta, it increases the internal optical
depth of the expanding ejecta, producing three observable phenomena:
(1) a mid-IR excess; (2) asymmetric blue-shifted emission
lines, since the dust obscures more emission from receding gas; and
(3) an increase in optical extinction.  All of these were observed
from one to three years after outburst in SN 1987A \cite{Luc91,Woo93}.
In this section, we present or confirm all three phenomena
from SN 2003gd.

\begin{figure}\centering
 \includegraphics[angle=0,width=0.9\linewidth]{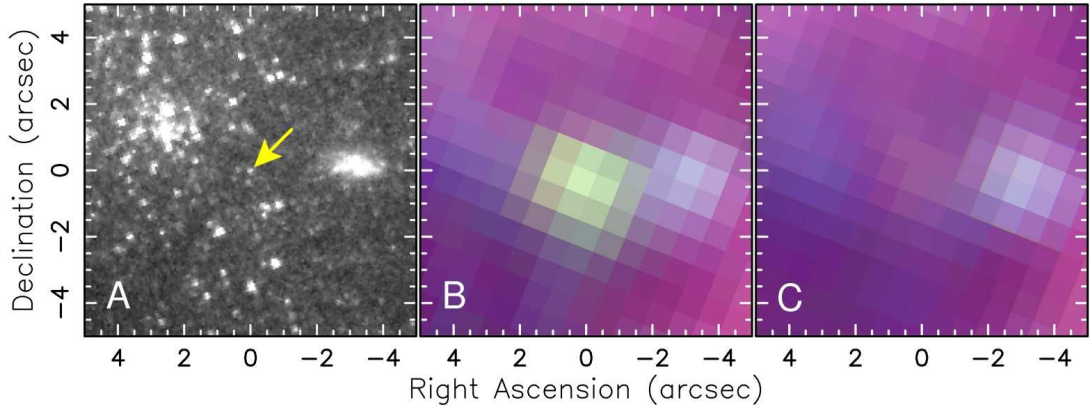}
\caption{Hubble and Spitzer Space Telescope images of a
  10\arcsec$\times$10\arcsec field centered on the position of SN
  2003gd.  ({\bf A}) HST WFPC2 image taken in the F606W filter in
  which the SN progenitor (arrowed) was identified \cite{VDy03,Hen05},
  with a resolution of 0\farcs045 pix$^{-1}$.  ({\bf B--C})
  False-color SST IRAC images of the SN taken 2004 Jul 28 ({\bf B})
  and 2005 Jan 15 ({\bf C}), showing the 3.6 \micron images in blue,
  4.5 \micron in green, and 8.0 \micron in red. All IRAC images
  were processed by the SINGS collaboration to a final resolution of
  0\farcs75 pix$^{-1}$.
\label{2003gdtile}}
\end{figure}

NGC 628 was observed by the Spitzer Infrared Nearby Galaxies Survey
(SINGS) Legacy program \cite{Ken03} with Spitzer Space Telescope's
Infrared Array Camera (IRAC) at 3.6, 4.5, 5.8, and 8.0 \micron on 2004
Jul 28, or day 499 after outburst \cite{Hen05}, and with the Multiband
Imaging Spectrometer for Spitzer (MIPS) on 2005 Jan 23 (day 678) at 24
\micron; the SN was also observed with IRAC as part of GO-3248 (P.I.\
W.\ P.\ Meikle) on 2005 Jan 15 (day 670).  All data were acquired from
the Spitzer archive, then spatially enhanced using the SINGS data
pipelines to a final resolution of 0\farcs75 pix$^{-1}$.  A point
source identified in all four IRAC bands from day 499 is consistent to
within 0\farcs17 (0.2 IRAC pixels) with the position of the SN
progenitor (Fig.\ \ref{2003gdtile}), as measured via careful absolute
and differential astrometry between the 3.6 \um IRAC image and the
archival Hubble Space Telescope (HST) Wide Field and Planetary Camera
2 (WFPC2) data in which the progenitor was identified \cite{Sma04}.

\begin{figure}\centering
 \includegraphics[angle=270,width=0.6\linewidth]{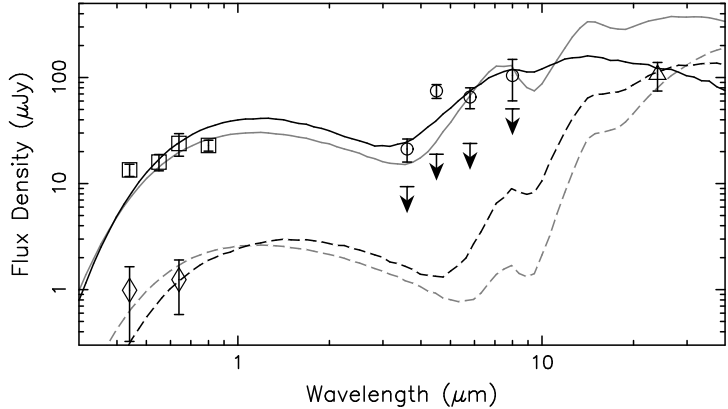}
\caption{Spectral-energy distribution of SN 2003gd, showing optical
  photometry on day 493 (squares) from \cite{Hen05} and extrapolated
  from day 632 \cite{Sug05} to day 678 (diamonds) using the evolution of
  SN 1987A \cite{SB90}; IRAC data from day 499 (circles); upper limits
  to IRAC from day 670 (arrows), and the MIPS datum from day 678
  (triangle).  Error bars are computed using a Poisson-noise model
  that includes detector charactertistics, flat-field and profile
  uncertainties.  Fluxes have been dereddened by $E(B-V)=0.14$
  \cite{Hen05}.  
  The curves are MOCASSIN radiative-transfer models to the data at
  day 499 (solid lines) and 678 (dashed lines) using
  smoothly-distributed (black) and clumpy (grey) dust.  See text and
  Table 2.
\label{fitbb}}
\end{figure}

Photometry of the Spitzer data was performed using
point-spread-function fitting techniques \cite{Ste87}.  The resulting
flux densities are $20.8\pm2.6$ \ujy at 3.6 \um, $73.8\pm5.6$ at 4.5
\um, $64.9\pm7.3$ \ujy at 5.8 \um, and $103\pm22$ \ujy at 8.0 \um on
day 499, and $106\pm16$ \ujy at 24 \um on day 678; the SN is nearly
undetectable in IRAC on day 670, with $3\sigma$ upper limits of 6.0,
10.6, 13.5, and 26.6 \ujy at 3.6, 4.5, 5.8, and 8.0 \um, respectively
(Fig.\ \ref{fitbb}).  An excess at 4.5 \micron could be due to CO
emitting at 4.6 \micron \cite{Woo93}.  Otherwise, the rising 5.8--24
\micron flux densities are not expected from the gaseous ejecta, which
typically have temperatures of $3000-5000$ K at these late times
\cite{Woo93}.  It is unlikely that this emission is a thermal light
echo, since the time variability of the IRAC fluxes is significantly
faster than that expected for a typical circumstellar dust shell
\cite{Dwe83}.  Assuming that the mid-IR emission can be modeled with a
single blackbody, the best fit to the 5.8--8.0 \um data from day 499
has a temperature of 480 K, which for an adopted distance to the SN of
9.3 Mpc \cite{Hen05}, yields an integrated luminosity of $4.6\times
10^5$ $L_\sun$ and an equivalent radius of $6.8\times 10^{15}$ cm.
The size, temperature, and variability implied by the spectral-energy
distributions (SEDs), are thus consistent with lower temperature ($T
\sim 500$ K) dust that is cooling within the SN ejecta.

\begin{figure}\centering
 \includegraphics[angle=270,width=0.6\linewidth]{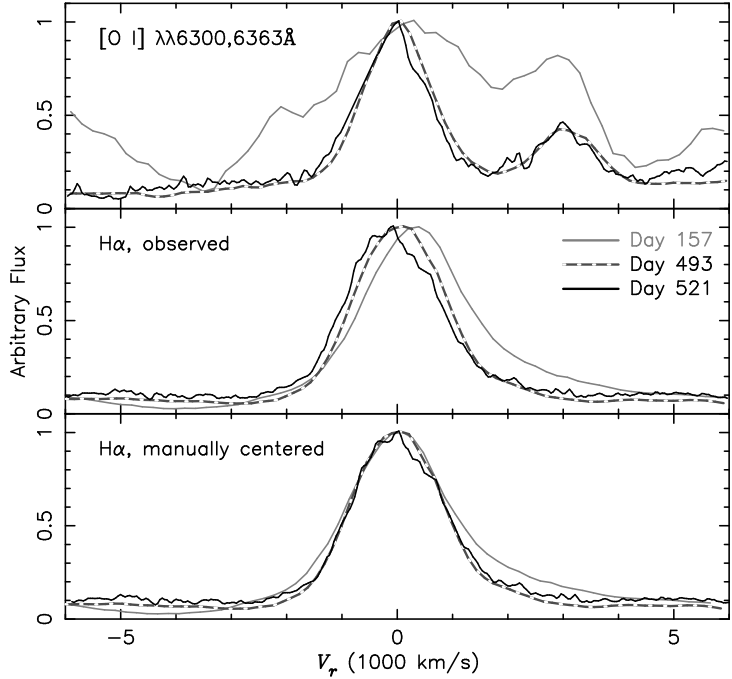}
\caption{Optical spectra of SN 2003gd showing [\ion{O}{I}]
  $\lambda\lambda$6300,6363 \AA\ (top panel) and H$\alpha$ (bottom
  panel).  Spectra have been corrected for the redshift $z=0.00219$ of
  NGC 628.  The grey curves are taken from the data presented in
  \cite{Hen05}, showing the profiles at days 157 (solid line) and 493
  (broken line).  The solid black curve is the new data taken on day
  521 with GMOS-N on Gemini North.  The profiles are normalized to an
  arbitrary flux scale.  A monotonic blueshift in the H$\alpha$ line
  peak, first reported by \cite{Hen05} is confirmed.  The most recent
  spectra in all three lines also show a clear profile asymmetry
  redward of each line peak.  This evolution is expected from dust
  forming homogenously within the ejecta, which preferentially
  extinguishes emission from the receding (i.e.\ redshifted) gas.
\label{Halpha}}
\end{figure}

The first indication of such dust formation in SN 2003gd came in a
comparison of broad-band photometry and H$\alpha$ spectra of the SN
between days 124 and 493 \cite{Hen05}, in which a small decline in the
late-time luminosity was accompanied by a slight blueshift in the
emission-line peaks.  New spectroscopic observations of SN 2003gd were
obtained in long-slit mode, covering $\sim$4500--7000 \AA\ with a
spectral resolution of $\sim$7 \AA\ using the Gemini Multi-Object
Spectrograph (GMOS) on Gemini North on 2004 August 19 (day 521). Two
spectra of SN 2003gd were obtained, which have been de-biased,
flattened, wavelength-calibrated, sky-subtracted, extracted and then
combined. The wavelength calibration is accurate to about 2 \AA.  The
H$\alpha$ and [\ion{O}{I}] spectra from days 157 and 493 \cite{Hen05}
are compared to the newer data in Fig.\ \ref{Halpha}.  Inspection of
the lines confirms that the emission peak is indeed blueshifting,
while the red high-velocity wing seen at the earliest epoch has
diminished.  Additionally, the most recent spectra show a clear
asymmetry in the first few hundred \kms redward of each line peak.
These blueshifted peaks and asymmetric profiles are consistent with a
simple model in which dust with an increasing optical depth is located
within an expanding sphere of uniform emission \cite{Elm03}; in
particular, the day 521 profiles suggest an optical extinction $A_R \lesssim
5$.

\begin{figure}\centering
 \includegraphics[angle=270,width=0.6\linewidth]{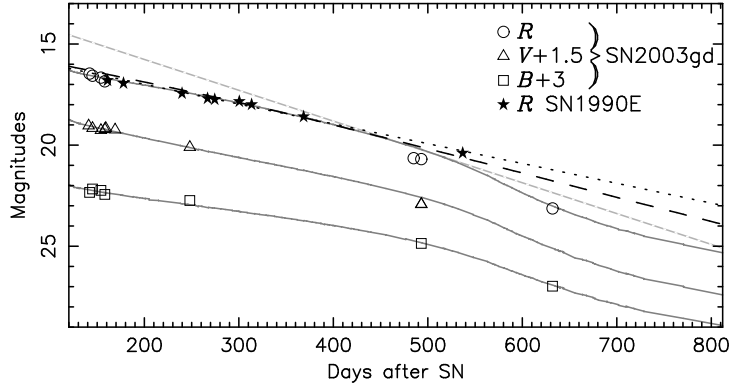}
\caption{Light curves of SN 2003gd, showing the increase of extinction
  with time.  $B$, $V$, and $R$ light curves of SN 2003gd compiled
  from \cite{Hen05} through day 493 after outburst, and \cite{Sug05}
  for day 632, are plotted and offset as marked.  Error bars are all
  smaller than the point markers.  For comparison, the corresponding
  light curves of SN 1987A \citep[][and references therein]{SB90} are
  also shown as thick grey lines, and the $R$-band light curve of SN
  1990E is shown as filled stars.  Also plotted in black are the light curves
  expected from Eq.\ (1) with (dashed) and without (dotted)
  the effective opacity term, as well as the linear fit to SN 1987A
  from days 450--525 (dashed grey) used by \cite{Luc91}.
\label{latelc}}
\end{figure}

The $B$, $V$, and $R$ light curves of SN 2003gd through day 493
\cite{Hen05} have been combined with the $B$ and $R$-band photometry
from day 632 \cite{Sug05} in Fig.\ \ref{latelc}.  Over this 500-day
period, the SN is evolving almost identically to SN 1987A, with the
exception that SN 2003gd is slightly subluminous in $V$ and $R$ around
day 493.  The close match between the light curves of these two SNe
implies an increase in optical extinction of SN 2003gd as well.

The observed dust extinction is measured by comparing the change in
photometry over time to a standard intrinsic luminosity.  The internal
extinction in SN 1987A is believed to have increased by $A_R=0.8$ mag
between days 525 and 700, as determined in broad-band photometry by
comparing the light curve after day 525 to the best-fit line through
the data from days 450--525 \cite{Luc91}.  The slope of this line
(Fig.\ \ref{latelc}) is extremely sensitive to the subset of the SN
1987A light curve used, and the resulting dust extinction can
vary by over 1 mag depending on the days included in the least-squares
fit.  Thus we deem this a poor measure of the intrinsic luminosity.

The broad-band evolution of Type II SNe past $\sim$500 days
is poorly documented, since very few SNe have ever been observed
beyond this epoch.  At these late times, the light curve is dominated
by the energy input of gamma rays from $^{56}$Co decay, which
decreases with an $e$-folding time of $\tau_{56}=111.2$ days.  The
$R$-band photometry of the Type II SN 1990E \cite{Ben94} closely
follows this evolution through 540 days (Fig.\ \ref{latelc}),
suggesting simple Co-decay is a good estimate of the unextinguished
$R$-band light curve for at least that long.   However, as the ejecta
expand, their opacity to gamma rays is expected to decrease, which
results in a modified light curve
\cite{Woo89}
\begin{equation}
 L_{56}(t) \propto e^{-t/\tau_{56}}\left[ 1-e^{ -\kappa_{56} \phi_0
 ({t_0}/{t})^2}\right]
\end{equation}
where the term in brackets is the ``effective opacity,'' with
$\kappa_{56}=0.033$~cm$^2$~g$^{-1}$ the average opacity to
$^{56}$Co-decay gamma rays, and $\phi_0=7\times10^4$~g~cm$^{-2}$ the
column depth at the fiducial time $t_0=11.6$ days, chosen to match the
bolometric light curve of SN 1987A \cite{Woo89,SB90}.  Note that Eq.\
(1) begins fading relative to simple $^{56}$Co decay around 500 days.

The Co-decay curves, both with and without the effective-opacity
correction, are much more luminous than the aforementioned linear fit
to SN 1987A at the time of dust formation (Fig.\ \ref{latelc}).  The
slopes of these Co-decay curves also closely resemble that of SN 1987A
after day 775, when dust production is believed to have ended
\cite{Woo93}.  Thus, Eq.\ (1) offers a more realistic standard
luminosity for the unextinguished $R$-band light curve of SN 1987A.
Comparison of $L_{56}$ to the SN 1987A photometry yields extinctions
of 1.5 mags around day 700 when effective opacity is included, and 2.5
mag when it is excluded.  Since the evolution of SNe 1987A and 2003gd
are so similar, Eq.\ (1) is also used to estimate the extinction of SN
2003gd at each epoch, as listed in Table 1.

\begin{table}\centering
\caption{$R$-band extinction of SN 2003gd. Eq.\ (1) is used with and
  without the effective opacity term to estimate the average and
  maximal extinction, respectively.  The observed values are listed
  first, and these were used along with the $R$-band light curve of SN
  1987A to extrapolate the extinction of the SN at the epochs of \SST
  observation.}
\begin{tabular}{l c c c}\hline
Day & \multicolumn{3}{c}{$A_R$ (mags)} \\
    &  average & maximal & error \\ \hline
\multicolumn{4}{c}{observed}\\
493 &  0.52 & 0.73  & 0.09 \\
632 &  1.36 & 1.78  & 0.12 \\ 
\multicolumn{4}{c}{extrapolated}\\
499 &  0.53 & 0.74  & 0.14 \\
678 &  1.51 & 2.13  & 0.20 \\ \hline
\end{tabular}
\end{table}

\section*{Dust-Mass Analysis and Interpretation}

The observations presented above overwhelmingly point toward dust
forming within the ejecta of SN 2003gd, beginning sometime between
250 and 493 days after outburst.  
To estimate the mass of dust present, we employ the three-dimensional
Monte Carlo radiative-transfer code MOCASSIN \cite{Erc05}.
Briefly, the paths of photon absorption, scattering and escape are
followed from a specified source through a given composition,
grain-size distribution, and geometry of dust.  The particular choices
of these are either constrained a priori or varied until the model
emission and extinction match the observed values.

Since our hypothesis is that dust condenses within the ejecta, the
radiative-transfer model is constructed under the initial assumption
that the dust and source luminosity are mixed within a spherical,
expanding shell with inner radius $r_{in}$, outer radius $Y\cdot
r_{in}$, and $\rho\propto r^{-2}$ density profile, and with the
illuminating radiation proportional to the dust density.  Initial
values for the shell size, source luminosity and temperature are
guided by the blackbody previously fit to the mid-IR data, and by
models of SN 1987A at similar epochs \cite{Woo93}. There are numerous
models for dust formation within SN ejecta \cite[for a review,
see][]{CN04}, most of which predict grain sizes will remain small.  We
adopt a standard $a^{-3.5}$ size distribution \cite{MRN77}, with grain
radii between $0.005-0.05$ \micron, and the dust composition is taken
to be 15\% amorphous carbon and 85\% silicates \cite{TF01}, with
optical constants taken from \cite{DL84,Han88}.  Finally, since there
are very few optical and mid-IR data to constrain a given model, the
source luminosity is restricted to evolve according to Eq.\ (1) while
its temperature remains constant \cite{Woo93}.

Two dust distributions are considered.  In the first, ``smooth''
model, the dust is uniformly distributed throughout the shell
according to the adopted density profile.  However, as early as a few
hours after outburst, post-shock ejecta become Rayleigh-Taylor
unstable \cite{Che78,HW94}, forming an inhomogenous or ``clumpy''
distribution, which we model as a two-phase medium, in which spherical
clumps with size $r_c=\delta\cdot(Y r_{in})$, volume filling factor
$f_c$ and density contrast $\alpha=\rho_c/\rho$ are embedded within an
interclump medium of density $\rho$.  This is analogous to the
mega-grains approximation \cite{VD99} with the addition of a radial
density profile.  Only macroscopic mixing has been found in the clumpy
ejecta of SN remnant Cas A \cite{DLC99}, which suggests elemental
ejecta layers remain heterogeneous.  We therefore assume the source
luminosity is completely separated from the dust clumps.  For a given
geometry, a clumpy model will always require more mass than a smooth
one to fit a given SED, since clumping lowers the overall optical
depth for a given mass of dust \cite{VD99}.  Rather than explore the
extensive parameter space of clumpy models, we study the limiting case
where all dust is in clumps, i.e. $\alpha\rightarrow\infty$, which
should provide upper mass limits, while the smooth models will provide
lower mass limits.  Finally, as suggested from hydrodynamic
simulations \cite{HW94}, we fix $\delta=0.025$.

\begin{table}\centering
\caption{Dust masses and $R$-band extinction calculated by the
   radiative-transfer code MOCASSIN.}
 \begin{tabular}{l c c c}
 \hline
 day\ \ \ \ & Model & $A_R$ & $M_{\rm dust}$ (\msun) \\ \hline
 499 & Smooth & 0.40 & $2.0\times10^{-4}$  \\
 499 & Clumpy & 0.65 & $1.7\times10^{-3}$  \\
 678 & Smooth & 1.48 & $2.7\times10^{-3}$  \\
 678 & Clumpy & 1.22 & $2.0\times10^{-2}$  \\ \hline
 \end{tabular}
\end{table}

Model results are summarized in Fig.\ \ref{fitbb} and Table 2.  A good
fit to the day 499 photometric and extinction data was achieved for
the smooth model using $Y=7$, $r_{in}=5\times10^{15}$ cm,
$L=6.6\times10^5$ \lsun, and $T=5000$ K, while fitting the day 678
data required changing $r_{in}$ to $6.8\times10^{15}$ cm and $L$ to
$9.2\times10^{4}$ \lsun.  Clumpy models used these same parameters,
with $f_c=0.02$ on day 499, and $f_c=0.05$ on day 678.  A complete
exploration of the model parameter space is beyond the scope of this
work, and will be presented elsewhere.  In general, small changes to
the model parameters have only modest effects.  For example, including
maximum grain sizes up to 0.25 \um (typical of dust in the
interstellar medium) decreases the dust mass by less than 10\%.  A
10\% change in $\delta$ or $f_c$ results in a 1--5\% change in mass
for our adopted parameter ranges.  Thus, the smooth and clumpy model
results shown in Table 2 offer reasonably robust lower and upper mass
limits, respectively.

These clumpy-model masses, up to $2\times 10^{-3}$ \msun on day 499
and $2\times 10^{-2}$ \msun on day 678, are significantly higher than
most analytic estimates of the dust mass for SN 2003gd.  For example,
using the same grain properties as above, $5\times10^{-4}$ and
$2\times10^{-3}$ \msun of dust are required to produce the mid-IR
emission at days 499 and 678, if all grains are visible and isothermal
\cite{Dot94}.  Using the mega-grains approximation for dust uniformly
mixed with diffuse emission within a spherical shell \cite{VD99}, the
$R$-band extinction yields masses of only $10^{-5}$ and
$4\times10^{-4}$ \msun of smooth dust for days 499 and 678.  In
contrast, up to $5\times10^{-3}$ \msun of clumpy dust is deduced
from the mega-grains model for day 499, which agrees well with our
radiative-transfer model.  However, once clumps become optically
thick, only geometry ($\delta,f_c$) determines their extinction, thus
an arbitrarily-large mass of clumpy dust reproduces the extinction
from day 678.  This behavior of the mega-grains model makes it of
limited use in determining dust masses when the observed optical depth
reaches unity.  We conclude that the most-often used analytic
approximations can provide unreliable estimates of dust masses.

Observations similar to those presented here have demonstrated the
condensation of dust in the ejecta of SN 1987A and 1999em, but the
inferred masses for these SNe were only modest, of order $10^{-4}$
\msun \cite{Woo93,Elm03}.  Asymmetric H$\alpha$ line profiles have
been detected for type II SNe 1970G, 1979C and 1980K \cite{Fes99},
however this phenomenon on its own is also consistent with an
expanding ionization front shell of emission catching up with and
passing through a pre-existing dust shell, and does not prove the
presence of newly-forming dust.  For SN 1998S, asymmetric line
profiles and near-IR ($\lambda<4.7$\um) excess emission also point
towards dust condensation in the ejecta \cite{Poz04}.  In all these
cases, the absence of contemporaneous mid-IR observations precluded a
quantitative estimate of the condensation efficiency.  In the light of
our quantitative clumpy-dust analysis, the amounts of dust believed to
have formed in the ejecta of SN 1987A and 1999em -- which were based
upon analytical estimates -- are being carefully revisited.

For a progenitor mass for SN 2003gd between 10--12 \msun \cite{Hen05},
roughly 0.16--0.42 \msun of refractory elements are expected to
form if the progenitor had solar metallicity \cite{Woo95}.  Assuming
dust formation has finished by day 678, our derived dust mass of
$2\times 10^{-2}$ \msun for this progenitor translates into a
condensation efficiency, defined here as (mass of refractory elements
condensed into dust)/(mass of refractory elements in ejecta), of $\le
0.12$ for clumpy dust.  Having found that analytic analyses of optical
and IR observations tend to underestimate the dust mass by an order of
magnitude or more, we deem it likely that dust formation in core
collapse SNe is significantly more efficient than previously believed.
In particular, the effiency implied by SN 2003gd is close to the value
of 0.2 needed for SNe to account for the dust content of high redshift
galaxies \cite{ME03}.  As noted earlier, too few SNe have been
followed sufficiently in time and wavelength to establish the
frequency with which SNe form dust.  We are currently addressing this
question through continued, long-term monitoring of a much larger
sample of young, type II SNe.  If dust formation is found to be common
in core collapse SNe, then since a majority of dust is expected to
survive later passage through high-velocity SN shocks \cite{Noz06}, we
can conclude that core collapse SNe played a significant role in the
production of dust in the early Universe.






\begin{scilastnote}
\item We gratefully acknowledge M.\ Hendry, for providing optical
  spectra of SN 2003gd, and E.\ Dwek and P. Ghavamian for useful
  discussions.  
  This work is based in part on archival data obtained with the 
  Spitzer Space Telescope, which is operated by the Jet Propulsion
  Laboratory, California Institute of Technology under a contract with
  NASA; 
  with the NASA/ESA Hubble Space Telescope, obtained from the
  Data Archive at the Space Telescope Science Institute, which is
  operated by the Association of Universities for Research in
  Astronomy, Inc. (AURA), under NASA contract NAS 5-26555;
  and on observations obtained during the program GN-2004B-C-3 at the
  Gemini Observatory, which is operated by AURA under a cooperative
  agreement with the NSF on behalf of the Gemini partnership.
  Support for B.E.K.S. for this work was provided by \SST award
  GO-20320 issued by JPL/Caltech.  D.L.W. acknowledges support from
  the Natural Sciences and Engineering Research Council of Canada
  (NSERC).  M.C.'s participation was supported by JPL contract
  \#1269553 with MIRA.
\end{scilastnote}

\end{document}